\documentstyle[aps]{revtex}

\begin{document}
\title{Off-diagonal structure of neutrino mass matrix in see--saw mechanism and
electron--muon--tau lepton universality}
\author{Riazuddin}
\address{National Center for Physics, Quaid-e-Azam University, Islamabad 45320,\\
Pakistan.}
\maketitle

\begin{abstract}
By a simple extension of the standard model in which ($e-\mu -\tau $)
universality is not conserved, we present a scenario within the framework of
see--saw mechanism in which the neutrino mass matrix is strictly
off--diagonal in the flavor basis. We show that a version of this scenario
can accomodate the small angle MSW solution of the solar neutrino problem
and $\nu _\mu -\nu _e$ oscillations claimed by the LSND collaboration.
Another version accomodates atmospheric $\nu _\mu -\nu _\tau $ oscillations
and large angle solution in solar neutrino experiments.

\end{abstract}

The minimal standard model involves three left-handed neutrino states and as
such does not admit renormalizable interactions that can generate neutrino
masses. However, there are three sets of data which suggest neutrino
oscillations, and hence that neutrinos have mass. One is thus forced to look
for viable extensions of the standard model. Before we consider one such
extension, let me summarize the three set of data \cite
{r1,r2,r3,r4,r5,r6,r7,r8} which claim neutrino oscillations:

$a)$ Atmospheric neutrnio experiments \cite{r4,r5,r6} [oscillations of $\nu
_\mu $ probabley into $\nu _\tau $] 
\begin{eqnarray*}
\Delta m_{\text{atm}}^2 &\equiv &\Delta m_{23}^2\simeq \left( 2-6\right)
\times 10^{-3}\text{ eV}^2 \\
\sin ^22\theta _{23} &\equiv &\sin ^22\theta _1\simeq 0.82-1.0
\end{eqnarray*}

$b)$ LSND collaboration for $\nu _\mu -\nu _e$ oscillations \cite{r7,r8} 
\begin{eqnarray*}
0.3\text{ eV}^2 &\leq &\Delta m_{12}^2\leq 2.0\text{ eV}^2 \\
10^{-3} &<&\sin ^22\theta <4\times 10^{-2}
\end{eqnarray*}

$c)$ Solar Neutrino Experiments \cite{r2,r3} 
\[
\begin{tabular}{ccccc}
MSW: &  & SMA: &  & $3\times 10^{-6}$ eV$^2\leq \Delta m_{23}^2\leq 10^{-5}$
eV$^2$ \\ 
&  &  &  & $2\times 10^{-3}<\sin ^22\theta _s<2\times 10^{-2}$%
\end{tabular}
\]
or 
\[
\begin{tabular}{ccccc}
MSW: &  & LMA: &  & $10^{-5}$ eV$^2\leq \Delta m_{12}^2\leq 10^{-4}$ eV$^2$
\\ 
&  &  &  & $0.6\leq \sin ^22\theta _3\leq 0.95$%
\end{tabular}
\]
or 
\[
\begin{tabular}{ccc}
V0: &  & $5\times 10^{-11}$ eV$^2\leq \Delta m_{12}^2\leq 5\times 10^{-10}$
eV$^2$ \\ 
&  & $0.6<\sin ^22\theta _3<1.0$%
\end{tabular}
\]
The three different mass splittings $\Delta m^2$ in (a), (b) and (c) above
seem do not compatible with a mixing of only three neutrino flavors.

It is known that there are two main mechanisms to generate tiny neutrino
masses \cite{altarelli}. One is the see-saw mechanism \cite{gellmann}
requiring the existance of superheavy $\left( \geq 10^{10}\text{ GeV}\right) 
$ right handed Majorana neutrinos while in the other tiny masses are \cite
{zee,frampton,matsuda} generated through higher order loop effects. Both
require an extension of the standard model and in both senarios light
neutrinos are Majorana.

A well known example of the second mechanism is the Zee model \cite{zee} in
which the neutrino mass matrix is strictly off-diagonal in the $\left( e,\mu
,\tau \right) $ basis. The purpose of this paper to revive a modest
extension of the standard model in which neutrinos have small masses and
lepton flavor [$\left( e,\mu ,\tau \right) $ universality] is not conserved.
It was shown \cite{fayyazuddin} that that this extension within the
framework of see--saw mechanism also lead to off-diagonal neutrino mass
matrix in flavor basis. One version of the Zee model can accomodate mass
hirarchy $\left| m_1\right| \simeq \left| m_2\right| \gg \left| m_3\right| ,$
$\Delta m_{13}^2\simeq \Delta m_{23}^2\simeq \Delta m_{\text{atm}}^2,\Delta
m_{12}^2\simeq \Delta m_{\text{solar}}^2,$ compatible with both atmospheric
and solar neutrino data, the latter has to be described by large angle
vacuum or MSW oscillations \cite{wolfenstein}. We show that in contrast a
version of our model accomodates mass heirarchy $\left| m_3\right| \simeq
\left| m_2\right| \gg \left| m_1\right| ,$ $\Delta m_{12}^2\simeq \Delta
m_{13}^2\simeq \Delta m_{\text{LSND}}^2$ and $\Delta m_{23}^2\simeq \Delta
m_{\text{solar}}^2$ for small angle MSW oscillations. However, there is
another version of our model where one gets the same results as in the Zee
model with $\left| m_1\right| \simeq \left| m_2\right| \gg \left| m_3\right|
,$ $\Delta m_{23}^2\simeq \Delta m_{13}^2\simeq \Delta m_{\text{atm}%
}^2,\Delta m_{12}^2\simeq \Delta m_{\text{solar}}^2.$

\section{Restrictions on neutrino mixing angles}

Let us consider an off-diagonal Majorana mass matrix in $\left( e,\mu ,\tau
\right) $ basis 
\begin{equation}
M=m_0\left( 
\begin{tabular}{ccc}
0 & $a_{e\mu }$ & $a_{e\tau }$ \\ 
$a_{e\mu }$ & 0 & $a_{\mu \tau }$ \\ 
$a_{e\tau }$ & $a_{\mu \tau }$ & 0
\end{tabular}
\right)  \label{01}
\end{equation}
It is convenient to define neutrino mixing angles as follows 
\begin{equation}
\left( 
\begin{tabular}{l}
$\nu _e$ \\ 
$\nu _\mu $ \\ 
$\nu _\tau $%
\end{tabular}
\right) =U\left( 
\begin{tabular}{l}
$\nu _1$ \\ 
$\nu _2$ \\ 
$\nu _3$%
\end{tabular}
\right)  \label{02}
\end{equation}
where 
\begin{equation}
U=\left( 
\begin{tabular}{ccc}
$c_2c_3$ & $c_2s_3$ & $s_2e^{i\delta }$ \\ 
$-c_1s_3-s_1s_2c_3e^{i\delta }$ & $c_1c_3-s_1s_2s_3e^{i\delta }$ & $s_1c_2$
\\ 
$s_1s_3-c_1s_2c_3e^{i\delta }$ & $-s_1c_3-c_1s_2s_3e^{i\delta }$ & $c_1c_2$%
\end{tabular}
\right)  \label{03}
\end{equation}
with $c_i=\cos \theta _i$ and $s_i=\sin \theta _i$. We shall put$\delta =0.$
Due to the off-diagonal structure of the mass matrix (\ref{01}), the
following relations are derived \cite{matsuda}: 
\begin{eqnarray}
m_2 &=&-\frac{\cos ^2\theta _3-\tan ^2\theta _2}{\sin ^2\theta _3-\tan
^2\theta _2}m_1,\text{ }m_1+m_2+m_3=0  \label{04} \\
\cos 2\theta _1\cos 2\theta _2\cos 2\theta _3 &=&\frac 12\sin 2\theta _1\sin
2\theta _2\sin 2\theta _3\left( 3\cos ^2\theta _2-2\right)  \label{05}
\end{eqnarray}

\begin{eqnarray}
-\sin 2\theta _1\cos 2\theta _2\cos 2\theta _3-\frac 12\cos 2\theta _1\sin
2\theta _2\sin 2\theta _3\left( 3\cos ^2\theta _2-2\right) &=&a_2a_{\mu \tau
}  \label{06} \\
-\cos 2\theta _3\sin \theta _1\sin \theta _2\cos \theta _2+\frac 12\cos
\theta _1\sin 2\theta _3\cos \theta _2\left( 3\cos ^2\theta _2-2\right)
&=&a_2a_{e\mu }  \label{07} \\
-\cos 2\theta _3\cos \theta _1\sin \theta _2\cos \theta _2-\frac 12\sin
\theta _1\cos \theta _2\sin 2\theta _3\left( 3\cos ^2\theta _2-2\right)
&=&a_2a_{e\tau }  \label{08}
\end{eqnarray}
where 
\begin{equation}
a_2=\frac{m_0}{m_2}\left( \cos ^2\theta _3\cos ^2\theta _2-\sin ^2\theta
_2\right) .  \label{n9}
\end{equation}
We also give here the transition probabilities 
\begin{eqnarray}
P\left( \nu _\mu -\nu _\tau \right) &=&\left[ -\right. \frac 14\sin
^22\theta _1\sin ^22\theta _3\left( 1+\sin ^2\theta _2\right) ^2+\sin
^22\theta _1\sin ^2\theta _2  \nonumber \\
&&+\cos 2\theta _1\sin 2\theta _1\sin 2\theta _3\cos 2\theta _3\sin \theta
_2\left( 1+\sin ^2\theta _2\right) \left. +\cos ^22\theta _1\sin ^22\theta
_3\sin ^2\theta _2\right] \sin ^2\left( \frac{\Delta m_{12}^2L}{4E}\right) 
\nonumber \\
&&+\sin 2\theta _1\cos ^2\theta _2\left[ \left( \sin 2\theta _1\cos ^2\theta
_3-\sin 2\theta _1\sin ^2\theta _2\sin ^2\theta _3+\sin 2\theta _3\cos
2\theta _1\sin \theta _2\right) \sin ^2\left( \frac{\Delta m_{23}^2L}{4E}%
\right) \right.  \nonumber \\
&&\left. +\left( \sin 2\theta _1\sin ^2\theta _3-\sin 2\theta _1\sin
^2\theta _2\cos ^2\theta _3-\sin 2\theta _3\cos 2\theta _1\sin \theta
_2\right) \sin ^2\left( \frac{\Delta m_{13}^2L}{4E}\right) \right]
\label{n10}
\end{eqnarray}
\begin{eqnarray}
P\left( \nu _e-\nu _\mu \right) &=&\left[ \sin ^22\theta _3\cos ^2\theta
_2\left( \cos ^2\theta _1-\sin ^2\theta _1\sin ^2\theta _2\right) +\sin
2\theta _1\sin 2\theta _3\sin \theta _2\cos ^2\theta _2\right] \sin ^2\left( 
\frac{\Delta m_{12}^2L}{4E}\right)  \nonumber \\
&&+\sin 2\theta _2\sin \theta _1\left[ \left( -\cos \theta _1\cos \theta
_2\sin 2\theta _3+\sin \theta _1\sin 2\theta _2\sin ^2\theta _3\right) \sin
^2\left( \frac{\Delta m_{23}^2L}{4E}\right) \right.  \nonumber \\
&&\left. +\left( \cos \theta _1\cos \theta _2\sin 2\theta _3+\sin \theta
_1\sin 2\theta _2\cos ^2\theta _3\right) \sin ^2\left( \frac{\Delta m_{13}^2L%
}{4E}\right) \right]  \label{n11}
\end{eqnarray}
\begin{eqnarray}
P\left( \nu _\mu -\nu _\tau \right) &=&\left[ \sin ^22\theta _3\cos ^2\theta
_2\left( \sin ^2\theta _1-\cos ^2\theta _1\sin ^2\theta _2\right) -\sin
2\theta _1\sin 2\theta _3\cos 2\theta _3\cos ^2\theta _2\sin \theta
_2\right] \sin ^2\left( \frac{\Delta m_{12}^2L}{4E}\right)  \nonumber \\
&&+\sin 2\theta _2\cos \theta _1\left[ \left( \sin 2\theta _3\sin \theta
_1\cos \theta _2+\cos \theta _1\sin ^2\theta _3\sin 2\theta _2\right) \sin
^2\left( \frac{\Delta m_{23}^2L}{4E}\right) \right.  \nonumber \\
&&\left. +\left( -\sin 2\theta _3\sin \theta _1\cos \theta _2+\cos \theta
_1\cos ^2\theta _3\sin 2\theta _2\right) \sin ^2\left( \frac{\Delta m_{13}^2L%
}{4E}\right) \right]  \label{n12}
\end{eqnarray}
\begin{equation}
P\left( \nu _e-\nu _e\right) =1-\cos ^4\theta _2\sin ^22\theta _3\sin
^2\left( \frac{\Delta m_{12}^2L}{4E}\right) -\sin ^22\theta _2\sin ^2\theta
_3\sin ^2\left( \frac{\Delta m_{23}^2L}{4E}\right) -\sin ^22\theta _2\cos
^2\theta _3\sin ^2\left( \frac{\Delta m_{13}^2L}{4E}\right)  \label{n13}
\end{equation}
It may be noted that 
\begin{equation}
\Delta m_{12}^2+\Delta m_{23}^2+\Delta
m_{31}^2=m_2^2-m_1^2+m_3^2-m_2^2+m_1^2-m_3^2=0  \label{n14}
\end{equation}

\section{Extension of the standard model and neutrino mass matrix}

By a simple extension of the standard electroweak gauge group to 
\[
G\equiv SU_L(2)\times U_e(1)\times U_\mu (1)\times U_\tau (1), 
\]
it was shown \cite{fayyazuddin} that the Majorana masses for light neutrinos
are generated through diagrams shown in figure 1.

Here $\phi ^{(i)}$ and $\Sigma ^{(i)}$ are respectively three $SU_L(2)$
Higgs doublets and singlets with appropriate $U_i(1)$ quantum numbers; $h$'s
and $f$'s are the corresponding Yukawa couplings. The symmetry is
spontaneously broken by giving vacuum expectation values to Higgs bosons $%
\phi ^{(i)}$ and $\Sigma ^{(i)}$: $\left\langle \phi ^{(i)}\right\rangle =%
\frac{v_i}{\sqrt{2}}$ and $\left\langle \Sigma ^{(i)}\right\rangle =\frac{%
\Lambda _i}{\sqrt{2}}$. For simplicity we shall take $v_1=v_2=v_3=v$ and $%
\Lambda _1=\Lambda _2=\Lambda _3=\Lambda $ (any difference can be absorbed
in the corresponding Yukawa couplings $h$ and $f)$. We take $\Lambda \gg v$
so that $X-$bosons which break the $e-\mu -\tau $ universality as well as
the the Majorana mass term for heavy neutrinos $N$'s are superheavy. In
order to simplify the calculation, we put $f_{12}=f_{13}=f_{23}=f$ (any
differences can again be absorbed in $h$-couplings) and put $f\Lambda /\sqrt{%
2}=M_R$. Thus finally we obtain the following off-diagonal mass matrix for
light neutrinos \cite{fayyazuddin} 
\begin{equation}
M_\nu =\frac{v^2}{2M_R}\left( 
\begin{tabular}{ccc}
$0$ & $h_1^{(2)}h_2^{(3)}$ & $h_1^{(2)}h_3^{(1)}$ \\ 
$h_1^{(2)}h_2^{(3)}$ & $0$ & $h_2^{(3)}h_3^{(1)}$ \\ 
$h_1^{(2)}h_3^{(1)}$ & $h_2^{(3)}h_3^{(1)}$ & $0$%
\end{tabular}
\right)  \label{09}
\end{equation}

The Yukawa couplings here are arbitrary and different choices for them
provide different predictions. We shall consider two choices, called $A$ and 
$B,$ with different mass hierarchies. For the choice $A$ we assume that the
Yukawa couplings are proportional to the generation index of quarks 
\begin{eqnarray}
h_1^{(2)}\frac v{\sqrt{2}} &=&\frac 1Km_u  \nonumber \\
h_2^{(3)}\frac v{\sqrt{2}} &=&\frac 1Km_c  \nonumber \\
h_3^{(1)}\frac v{\sqrt{2}} &=&\frac 1Km_t  \label{10}
\end{eqnarray}
where K is dimensioless parameter. Further we take \cite{altarelli} $%
m_u:m_c:m_t=\lambda ^6:\lambda ^4:1$ as an order of magnitude relations.
Then the mass matrix (\ref{09}) can be written as 
\begin{equation}
M_\nu =m_0\left( 
\begin{tabular}{ccc}
0 & $\lambda ^6$ & $\lambda ^2$ \\ 
$\lambda ^6$ & 0 & 1 \\ 
$\lambda ^2$ & 1 & 0
\end{tabular}
\right)   \label{11}
\end{equation}
where 
\begin{equation}
m_0=\frac{\lambda ^4m_t^2}{K^2M_R}.  \label{new12}
\end{equation}
In the first approximation, it has eigenvalues $m_0\left( \pm 1,0\right)
\left[ m_2\simeq -m_3\right] $. The right-hand sides of Eqs. (\ref{06}), (%
\ref{07}) and (\ref{08}) respectively become 1, $\lambda ^6$ and $\lambda ^2.
$ Eqs. (\ref{07}) and (\ref{08}) can then only be satisfied if both $\theta
_2$ and $\theta _3$ are small so that $\sin \theta _{2,3}\simeq \theta _{2,3}
$ while Eqs. (\ref{05}) and (\ref{06}) are also then satisfied for $\sin
2\theta _1\simeq 1,$ $\cos 2\theta _1\simeq 0,$ $\sin \theta _1\simeq 1/%
\sqrt{2},$ $\cos \theta _1\simeq 1/\sqrt{2}.$ Writing Eqs. (\ref{06}) and (%
\ref{07}) in detail we have then 
\begin{eqnarray}
-\frac{\theta _2}{\sqrt{2}}+\frac 12\frac 1{\sqrt{2}}\sin 2\theta _3
&=&-\lambda ^6  \nonumber \\
-\frac{\theta _2}{\sqrt{2}}-\frac 12\frac 1{\sqrt{2}}\sin 2\theta _3
&=&-\lambda ^2  \label{12}
\end{eqnarray}
implying 
\begin{eqnarray}
\sin 2\theta _3 &\simeq &\sqrt{2}\lambda ^2\left( 1-\lambda ^4\right) 
\label{13} \\
\sin 2\theta _2 &\simeq &2\theta _2=\sqrt{2}\lambda ^2\left( 1+\lambda
^4\right)   \label{14}
\end{eqnarray}
Finally from Eq. (\ref{04}), $m_2\simeq -1,$ $m_1\simeq 0$ so that $%
m_3\simeq 1$. In fact diagonalization of the matrix (\ref{11}) give 
\begin{eqnarray}
m_3 &\approx &m_0\left( \sqrt{1+\lambda ^4}+\lambda ^8\right)   \nonumber \\
m_2 &\approx &m_0\left( -\sqrt{1+\lambda ^4}+\lambda ^8\right)   \nonumber \\
m_1 &\approx &-2\lambda ^8m_0  \label{15}
\end{eqnarray}
so that $m_3\simeq \left| m_2\right| \gg \left| m_1\right| $ and 
\begin{equation}
\Delta m_{12}^2=\Delta m_{13}^2=m_0^2,\quad \Delta m_{23}^2\simeq
4m_0^2\lambda ^8.  \label{n23}
\end{equation}
Finally from Eqs. (\ref{02}) and (\ref{03}) to leading orders in $s_2$ and $%
s_3$%
\begin{eqnarray}
\nu _1 &\simeq &\nu _e-s_3\left( c_1\nu _\mu -s_1\nu _\tau \right)
-s_2\left( s_1\nu _\mu +c_1\nu _\tau \right)   \nonumber \\
\nu _2 &\simeq &s_3\nu _e+c_3\left( c_1\nu _\mu -s_1\nu _\tau \right)  
\nonumber \\
\nu _3 &\simeq &s_2\nu _e+c_2\left( s_1\nu _\mu +c_1\nu _\tau \right) 
\label{16}
\end{eqnarray}
showing that $\nu _1$ is primarily $\nu _e$ while $\nu _2$ and $\nu _3$ are
primarily $\left( c_1\nu _\mu -s_1\nu _\tau \right) $ and $\left( s_1\nu
_\mu +c_1\nu _\tau \right) $ respectively.

We now consider the choice $B$, where we assume $h_1^{(2)}\gg
h_1^{(3)}\simeq h_2^{(3)}$ and use the parametrization 
\begin{eqnarray}
h_2^{(3)} &=&h\cos \theta ,\quad h_1^{(3)}=h\sin \theta  \nonumber \\
\frac{h_1^{(3)}h_2^{(3)}}{h_1^{(2)}} &=&h\sigma ,\quad m_0=\frac{%
hh_1^{(2)}v^2}{2M_R}
\end{eqnarray}
where $\sigma \ll 1.$ Then 
\begin{equation}
M_\nu =m_0\left( 
\begin{tabular}{ccc}
0 & $\cos \theta $ & $\sin \theta $ \\ 
$\cos \theta $ & 0 & $\sigma $ \\ 
$\sin \theta $ & $\sigma $ & 0
\end{tabular}
\right)
\end{equation}
The diagonalization (neglecting $\sigma ^2$) gives 
\begin{eqnarray}
m_{2,1} &=&m_0\left[ \pm 1+\frac 12\sigma \sin 2\theta \right]  \nonumber \\
m_3 &=&-\left( m_1+m_2\right) =-m_0\sigma \sin 2\theta
\end{eqnarray}
so that $\left| m_1\right| \simeq \left| m_2\right| \gg \left| m_3\right| $
and 
\begin{eqnarray}
\Delta m_{12}^2=2\sigma \sin 2\theta ,\quad \Delta m_{31}^2=\Delta
m_{32}^2=m_0^2
\end{eqnarray}
We will take $\theta _2\simeq 0$ as before so that from Eq. (\ref{04}), we
must have $\theta _3\simeq \frac \pi 4$ in order to have $m_2=-m_1.$ Then
from Eqs. (\ref{06}), (\ref{07}) and (\ref{08}) in leading order, $\theta
=-\theta _1.$ In this case 
\begin{eqnarray}
\nu _1 &\simeq &\frac 1{\sqrt{2}}\left[ \nu _e-\left( \cos \theta _1\nu _\mu
-\sin \theta _1\nu _\tau \right) \right]  \nonumber \\
\nu _2 &\simeq &\frac 1{\sqrt{2}}\left[ \nu _e+\left( \cos \theta _1\nu _\mu
-\sin \theta _1\nu _\tau \right) \right]  \nonumber \\
\nu _3 &\simeq &\left( \sin \theta _1\nu _\mu +\cos \theta _1\nu _\tau
\right)  \label{n29}
\end{eqnarray}

\section{Transition probabilities and conclusions}

For our choice $A$, $\lambda $ is expected to be of order $0.22\simeq \sin
\theta _c$ ($\theta _c$ being the Cabibbo angle) so that from Eqs. (\ref{13}%
), (\ref{14}) and (\ref{n23}) 
\begin{eqnarray}
\sin ^22\theta _3 &\simeq &\sin ^22\theta _2\simeq 2\lambda ^4\simeq
4.5\times 10^{-3}  \label{n30} \\
\Delta m_{23}^2 &=&2\times 10^{-5}m_0^2  \label{n31}
\end{eqnarray}
Thus with 
\begin{equation}
\Delta m_{12}^2\simeq \Delta m_{31}^2\gg \Delta m_{23}^2  \label{18}
\end{equation}
and neglecting terms of order $s_3^4,s_2^2s_3^2,\cos 2\theta _1s_2s_3,$ we
have from Eqs. (\ref{n10})--(\ref{n12}) 
\begin{eqnarray}
\left. P\left( \nu _\mu -\nu _\tau \right) \right| _{\text{atm}} &=&\sin
^22\theta _1\cos ^2\theta _2\cos ^2\theta _3\sin ^2\frac{\Delta m_{23}^2R_a}{%
4E}  \nonumber \\
&\simeq &\sin ^22\theta _1\sin ^2\frac{\Delta m_{23}^2R_a}{4E}  \label{19}
\end{eqnarray}
\begin{equation}
\left. P\left( \nu _e-\nu _\mu \right) \right| _{\text{LSND}}=\left[ \sin
^22\theta _3\cos ^2\theta _1+\sin 2\theta _1\sin \theta _2\sin 2\theta
_3+\sin 2\theta _2\sin 2\theta _3\sin \theta _1\cos \theta _1+\sin ^22\theta
_2\sin ^2\theta _1\right] \sin ^2\frac{\Delta m_{12}^2R_{\text{LSND}}}{4E}
\label{20}
\end{equation}
and 
\begin{eqnarray}
P\left( \nu _e-\nu _\tau \right)  &=&\left[ \sin ^2\theta _1\sin ^22\theta
_3-\sin 2\theta _1\sin \theta _2\sin 2\theta _3-\sin \theta _1\cos \theta
_1\sin 2\theta _2\sin 2\theta _3+\sin ^22\theta _2\cos ^2\theta _1\right]
\sin ^2\frac{\Delta m_{12}^2R_s}{4E}  \nonumber \\
&&+\left[ \sin \theta _1\cos \theta _1\sin 2\theta _2\sin 2\theta _3\right]
\sin ^2\frac{\Delta m_{23}^2R_s}{4E}  \label{n35}
\end{eqnarray}
Now with $\theta _1\simeq \frac \pi 4$ and using Eq. (\ref{n31}) [note that
the coefficient of $\sin ^2\left( \Delta m_{12}^2R_s/4E\right) $ in Eq. (\ref
{n35}) vanishes], we have from Eqs. (\ref{20}) and (\ref{n35}) 
\begin{eqnarray}
\left. P\left( \nu _e\rightarrow \nu _\mu \right) \right| _{\text{LSND}}
&\simeq &\sin ^22\theta _{\text{eff}}\sin ^2\frac{\Delta m_{12}^2R_{\text{%
LSND}}}{4E}  \label{24} \\
\left. P\left( \nu _e\rightarrow \nu _\tau \right) \right| _{\text{solar}}
&\simeq &\sin ^22\tilde{\theta}_{\text{eff}}\sin ^2\frac{\Delta m_{23}^2R_s}{%
4E}  \label{n37}
\end{eqnarray}
where 
\begin{eqnarray}
\sin ^22\theta _{\text{eff}} &\simeq &2\left( 4.5\right) \times
10^{-3}\simeq 10^{-2}  \label{25} \\
\sin ^22\tilde{\theta}_{\text{eff}} &\simeq &\frac 12\left( 4.5\right)
\times 10^{-3}=2.25\times 10^{-3}  \label{n39}
\end{eqnarray}
Thus with $m_0^2\simeq 0.3$ eV$^2,$ $\Delta m_{23}\simeq 4\lambda
^8m_0^2\simeq 2\times 10^{-5}m_0^2\simeq 6\times 10^{-6}$ eV$^2$ the LSND
data and solar neutrino oscillations are explained, the latter with SMA MSW
solution. Finally with $m_0\simeq \sqrt{0.3}$ eV, $\lambda ^4=4.5\times
10^{-3}$ and $m_t\simeq 175$ GeV, we obtain from Eq. (\ref{new12}) $%
K^2M_R\simeq 10^{11}$ GeV, giving the mass scale at which $e-\mu -\tau $
universality is broken and the scale associated with superheavy Majorana
neutrinos. In the version of the model we have considered $L_\tau -L_\mu -L_e
$ number is , however, conserved while in the particular version of the Zee
model mentioned earlier as well as in our model $B$, it is the $L_e-L_\mu
-L_\tau $ number which is conserved.

We now consider the predictions of our version $B$ for which 
\begin{equation}
m_0^2=\Delta m_{31}^2=\Delta m_{32}^2\gg \Delta m_{12}^2  \label{n40}
\end{equation}
and $\theta _3\simeq \frac \pi 4,$ $\theta _2\simeq 0$ so that neglecting $%
\sin ^22\theta _2$ and $\cos 2\theta _3,$ we obtain in the leading order
from Eqs. (\ref{n10}) and (\ref{n13}) 
\begin{eqnarray}
\left. P\left( \nu _e\rightarrow \nu _\tau \right) \right| _{\text{atm}}
&\simeq &\sin ^22\theta _1\sin ^2\left( \frac{\Delta m_{32}^2R_a}{4E}\right)
\label{n41} \\
\left. P\left( \nu _e\rightarrow \nu _e\right) \right| _{\text{solar}}
&\simeq &1-\sin ^22\theta _3\sin ^2\left( \frac{\Delta m_{12}^2R_s}{4E}%
\right)  \label{n42}
\end{eqnarray}
Thus atmospheric neutrino experimental data is explained with $\Delta
m_{32}^2\simeq m_0^2\simeq 10^{-3}$ eV$^2$ and $\theta _1\simeq \frac \pi 4$
while the Eq. (\ref{n42}) is consistent with the large angle [$\sin
^22\theta _3\simeq 1$] vacuum or MSW solution in solar neutrino experiments.
Here $m_3,$ the mass of the lightest neutrino consisting mainly of $\nu _\mu 
$ and $\nu _\tau $ [cf. Eq. (\ref{n29})] is given by 
\[
m_3\simeq \sigma \sin 2\theta _1m_0=\sigma m_0=\frac{\Delta m_{12}^2}{2m_0}=%
\frac{\Delta m_{\text{solar}}^2}{2\sqrt{\Delta m_{\text{atm}}^2}}. 
\]

To conclude by considering a simple extention of the standard model in which 
$\left( e-\mu -\tau \right) $ universality is not conserved, we have
presented a scenario within the framework of see--saw mechanism in which the
neutrino mass matrix is strictly off-diagonal in the flavor basis. Further
we have shown that a version of this scenario can accomodate the atmospheric 
$\nu _\mu -\nu _\tau $ neutrino oscillations and large angle vacuum or MSW
solution in solar neutrino experiments while another version is compatible
with small angle MSW solution of solar neutrino oscillations and $\nu _\mu
-\nu _e$ oscillations claimed by the LSND collaboration.

\end{document}